\documentstyle[art10,epsf]{article}
\begin{document}

{\Large\bf
An Inconsistency in the Simulation of \\
Bose-Einstein Correlations
}\\[3mm] 
\def\rightmark{An Inconsistency in the Simulation of Bose-Einstein 
Correlations}\def\leftmark{M. Martin et al.}
\hspace*{6.327mm}\begin{minipage}[t]{12.0213cm}{\large\lineskip .75em
M. Martin$^1$, 
H. Kalechofsky$^1$, 
P. Foka$^1$,
and U.A. Wiedemann$^2$
}\\[2.812mm] 
\hspace*{-8pt}
$^1$ University of Geneva, Geneva, Switzerland, \\
\hspace*{-8pt}
$^2$ Institut f\"ur Theoretische Physik, Universit\"at Regensburg, D-93040
Regensburg, Germany \\[0.2ex]
\\[4.218mm]{\it
Received nn Month Year (to be given by the editors)
}\\[5.624mm]\noindent
%
%
%
{\bf Abstract.}
We show that the formalism commonly used to
implement Bose-Einstein correlations in Monte-Carlo simulations can lead
to values of the two-particle correlator significantly smaller than unity,
in the case of sources
with strong position-momentum correlations.
This is more pronounced when the phase space of the emitted particles 
is strongly reduced by experimental acceptance or kinematic 
analysis selections. It is inconsistent with general principles 
according to which the Bose-Einstein correlator is larger than unity.
This inconsistency seems to be rooted in the fact that 
quantum mechanical localization properties
are not taken into account properly.
\end{minipage}

\section{Introduction}

 Hanbury-Brown Twiss interferometry has been used \cite{Gol1,Han1} in 
both high energy and nuclear physics to determine the space-time 
dimensions of the emitting source created during nuclear collisions
by using the effect of the interference pattern between two identical 
produced bosons \cite{Sey1,Bek1,Abb1}. 

 The source parameters derived from fits
to the correlation function are difficult to interpret directly
as real geometric quantities, being sensitive to the transverse and
longitudinal dynamical expansion of the system 
\cite{Her1,Cso1,Cha1,Cha2,Akk1,Alb1}, which result 
in a momentum dependence of the extracted source 
radii \cite{Wie1,Bek2,Aki1} to long lived resonance decays 
\cite{Pad1,Sul1,Cso2,Wie2,Sch1},
the Coulomb interaction \cite{Pra1,Bay1}, and
final state rescattering \cite{Hum1}. There has been recent experimental 
evidence for flow effects \cite{Xu1,Alb1} and one possible
implication of flow 
is that distant points of emission in the source volume cannot emit  
particles with closely differing  momenta, and thus do not contribute to 
the small relative pair momentum region \cite{Mak1}.  
It is also anticipated that strong absorption must exist in the case of 
large stopping power; a particle originating at the side of the source
opposite to the direction of its momentum cannot easily propagate through 
the source to be seen by the detector, and therefore only
a limited region of the source will be seen, noted already
in AGS studies \cite{Nay1}. It is thus interesting to try to further
probe the relationship between source geometry, dynamical expansion, 
and kinematical regimes viewed by the measuring apparatus. 

To study the effect of position-momentum correlations 
on the shape of the correlation functions, a simple Monte Carlo 
phase-space model controlled by a few macroscopic parameters was developed.
As is the case with more detailed and sophisticated microscopic event 
generators, there is no Bose-Einstein symmetrization effect included from
first principles \cite{Wer1,Sor1,Hum1}. The
Bose-Einstein correlations were then added to the initial
distributions by including the symmetrization in the form
of a weight calculated for each pair of identical particles,
a procedure found extensively in the literature 
\cite{Pra2,Sul1,Hum1,Fie1}.

In the present work we show that there are limitations
to this formalism, as it is an approximation and several assumptions
are implicit in its derivation.
This will be demonstrated on the
basis of the Monte-Carlo calculations. 
We briefly check that our model produces results
consistent with theoretical predictions as well as 
with other simulation studies
in case of similar phase space distributions.
In our further studies of source
models with position momentum dependence, 
we then find that the current praxis
of including Bose-Einstein correlations is  
inconsistent with general principles. 
Hence we restrict our discussion to pointing to a
possible origin of this problem.
 
This paper proceeds as follows: Section 2 describes the
Monte-Carlo model and the inclusion of the Bose-Einstein 
effect, Section 3 presents the model results and
in Section 4, we turn to the discussion of the 
observed problems.  

\section{The Model}

 Our model produces events from an analytically given phase space
distribution by a Monte Carlo technique. The model provides the 
phase space distribution of particles at the points 
($x^{\mu},p^{\mu}$) of their last interaction, with no assumptions
about the dynamical evolution of the collision. 
The collision region is described in terms of a few 
macroscopic parameters defining the spatiotemporal extension of the
source, such as the source shape and size, and the dynamical
features of the system, such as temperature and collective flow.
All the particles are assumed to be pions, and resonances are 
not included. 

As for all existing event generators, the obtained phase space 
distribution does not contain Bose-Einstein correlations. To
include the latter, we applied the following prescription, widely
found and used in the literature ~\cite{Pra1,Hum2}. Each identical
pion pair emitted from the points 
$({\vec{r_1}},t_1)$ and $({\vec{r_2}},t_2)$ is weighted
with the Born probability 
of a symmetrized two-pion plane wave, 
  \begin{equation}
    \mid\Psi\mid^2 = 1 + cos [({\vec{r_1}}-{\vec{r_2}}) 
    \cdot ({\vec{p_1}}-{\vec{p_2}})
    -(t_1-t_2)(E_1-E_2)] 
    \label{5}
  \end{equation}
which reduces to
  \begin{equation}
    \mid\Psi\mid^2 = 1 + cos [({\vec{r_1}}-{\vec{r_2}}) 
    \cdot ({\vec{p_1}}-{\vec{p_2}})]
    \label{6}
  \end{equation}
since the pions at freezeout are on-shell by definition and 
the source emission is assumed to be instantaneous (there is no time 
evolution).
The correlator defined via Eq. (\ref{6}) is a function of the 
three-dimensional relative momentum component ${\bf q}$. 
 The correlator $C({\bf q})$ is then obtained for each bin 
as the sum over pion pairs weighted by $\mid\Psi\mid^2$ and normalized 
to the sum of unweighted pion pairs, cf.~\cite{Pra1}. The 
Coulomb interaction between the pions is not simulated.

Here, we streamline our presentation by restricting it to 
the results of the one-dimensional fits of $C(q_3)$ in terms 
of the 3-momentum difference  $q_{3}$
  \begin{equation}
    C(q_{3})=1+\lambda \mbox{exp} [-q^2_{3}R^2_{3}]\, 
    \qquad \qquad q_{3} = \mid \vec{p_1} - \vec{p_2} \mid \, .
    \label{7}
  \end{equation}
All calculations are 
done in the longitudinal comoving system {\em LCMS} frame,
which is defined such that the longitudinal component of the
average pair momentum vanishes. 

\section{Model Results}

In contrast to full event generators like Venus, RQMD, ARC, etc.,
which try to incorporate all the physics expected to be present
in a heavy ion collision, the purpose of our model was to
isolate and study one important effect: the geometrical and
dynamical interpretation of HBT parameters in the presence of 
radial flow and realistic experimental acceptance.
The model's simplicity allowed the well controlled study of a 
wide set of different flow and acceptance conditions. 
Thus in the course of this study, we have found, inadequacies in
the common practice (Eq. (\ref{5})) of including Bose-Einstein
correlations which become particularly apparent
for sources with strong position-momentum dependence
where certain kinematical selections are imposed. 
Our presentation will first illustrate this effect
in an instructive way for the extreme case of complete 
position-momentum correlations in the source. Next,
we will summarize the results of flow and acceptance effects
on the HBT parameters for realistic phase space distributions
and acceptance criteria.

\subsection{Source with Complete Position-Momentum Correlations}
\label{sec3.1}

\begin{figure}[p]
 \begin{center}\mbox{\epsfxsize=14cm\epsffile{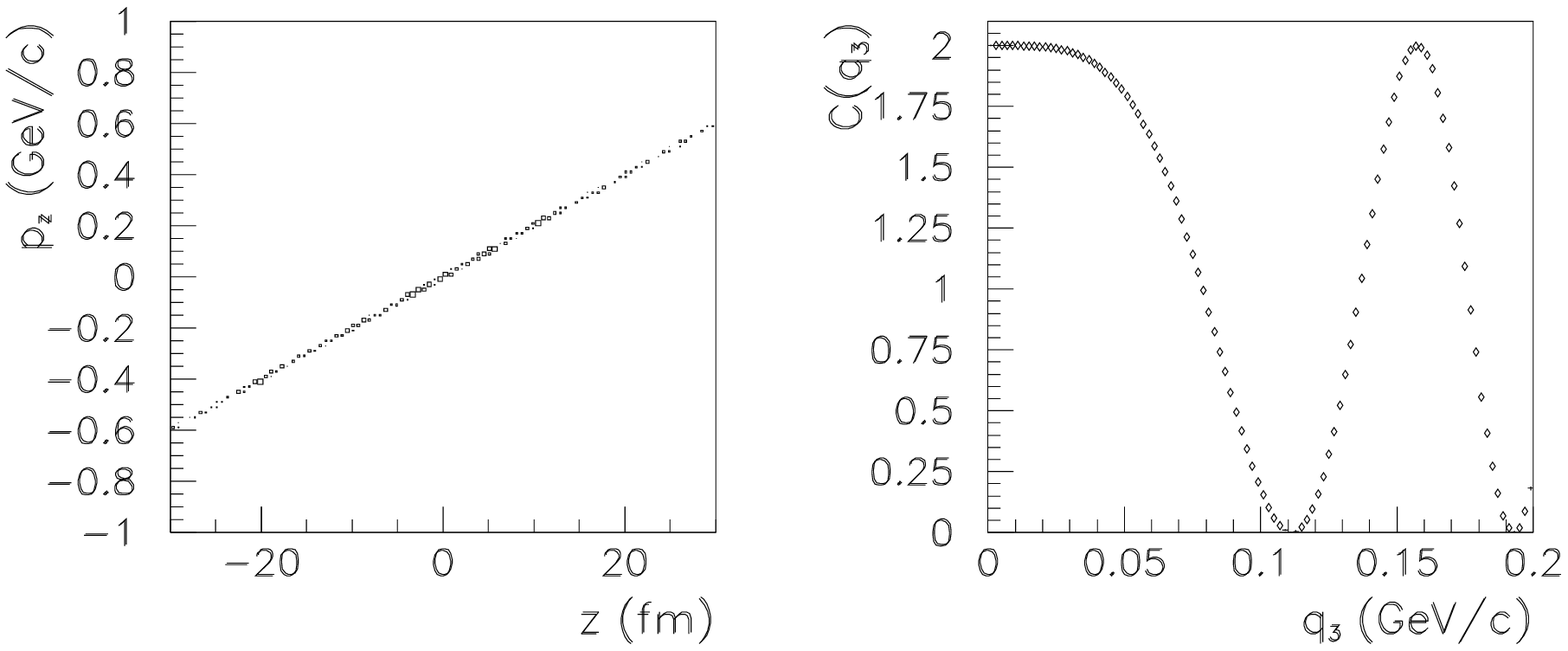}}\end{center}
 \caption{For a linear source expanding in the $z$ direction according
to Eq. (\ref{8}), $D = 0.02$. 
{\bf Left}: Momentum $p_z$ in the beam direction as
a function of the $z$ position of emitted pions. {\bf Right}: The resulting
two-pion correlation in $q_{3}$ using the formula in Eq. (\ref{6}).}
 \label{zpz1}
\end{figure}

Here, we calculate the correlation function for a linear 
source in the beam ($z$) direction with neither 
transverse spatial extension nor transverse momentum dependence.
The longitudinal momentum is chosen to be completely due to flow, 
  \begin{equation}
    p(z) = D\, z\, ,
    \label{8} 
  \end{equation}
where $p$ has units of GeV/c, $D$ being a constant. This distribution
represents a source expanding in the $z$-direction for which the
argument in the cosine of Eq. (\ref{6}) reduces to 
$({\vec{r_1}}-{\vec{r_2}}) \cdot 
({\vec{p_1}}-{\vec{p_2}})= Q_{3}^{2}/{D^2} $. For $D = 0.02$,
this leads to the correlation function shown in Figure ~\ref{zpz1}. 
The source in Eq. (\ref{8}) has a total position-momentum correlation
and in this case,
as can be checked analytically \cite{Hei2},
the correlator obtained 
for Eq. (\ref{8}) with the $cos$-prescription oscillates 
between 2 and 0. Introducing a Gaussian spatial smearing 
of the emission points in Eq. (\ref{8}), as might be motivated 
by the picture of a limited quantum mechanical localization 
of particles, one sees that the oscillations of the 
correlator decrease. Still, the correlator can drop below unity, and this
is in strong contradiction to calculations from first principles
\cite{Hei1} which ascertain that for arbitrary sources the correlator 
is always larger than unity. We next investigate in how far this
behaviour persists for more realistic phase space distributions.

\subsection{Position-Momentum Dependence in a Realistic Model}

To study more realistic scenarios, we generate pions
according to macroscopic model parameters
that correspond to an instantaneous Gaussian source with
realistic momentum distributions.
No attempt is made either to reproduce any data or to create a 
fully realistic Monte-Carlo event generator.
Accordingly the spatiotemporal part of the source is 
modelled by 
  \begin{equation}
    G = {\rm const.} \times \exp{[-(x^2+y^2+z^2)/R^2]} \, .
    \label{1}
  \end{equation}
The transverse momentum dependence and the rapidity dependence chosen
for the case of no flow (i.e., no position momentum dependence in the
source) are
  \begin{eqnarray}
    dN/dp_T &=& A\, p_T \exp{(-p_T/B)}\, ,
    \nonumber \\
    y &=& c\sqrt{-2 \ln a}(cos(2\pi b))\, .
    \label{2}
  \end{eqnarray}
As an input parameter, we use a Gaussian radius
$R = 6$ fm, motivated by the hard sphere radius of the $^{208}Pb$  
incoming projectile. 
The input for the momentum distributions is chosen according to
the measured transverse momentum ($p_T$) and rapidity ($y$) 
distributions in the CERN 158 GeV/n Pb+Pb data \cite{qm95}.
The inverse slope parameter of the $p_T$-distribution
is taken to be $B = 200$ MeV and $A$ is an arbitrary normalization 
constant. The rapidity ($y \equiv  \frac{1}{2} \log{\frac{E+p_z}{E-p_z}}$) 
dependence is specified by random numbers $a$ and $b$ which are 
uniformly distributed in the interval (0,1), $c$ being a constant.

\begin{figure}[p]
 \begin{center}\mbox{\epsfysize=10cm\epsffile{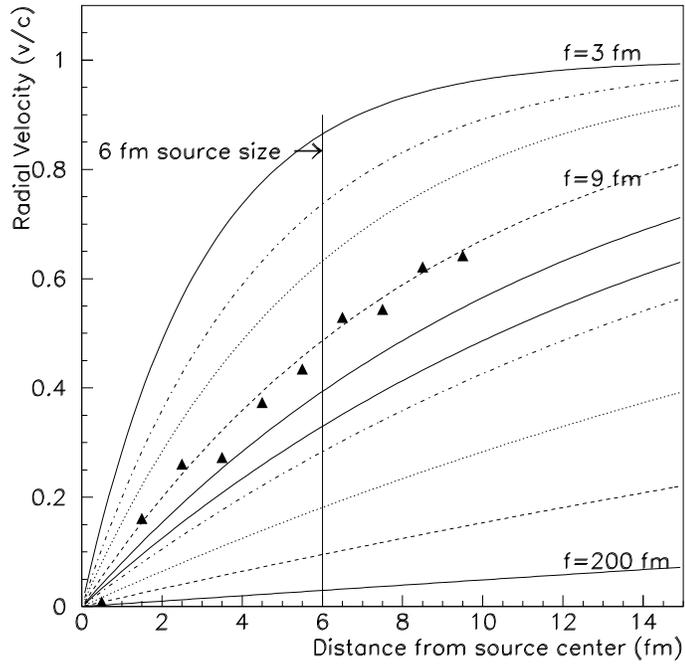}}\end{center}
 \caption{The radial flow velocity as a function of the radial distance 
from the center of a pure Gaussian source parametrized according 
to Eq. (\ref{7}). The flow extracted from Venus (v. 4.12) is well 
represented by this parametrization with $f$=9 fm (filled triangles). 
The vertical line shows the true 6 fm Gaussian source size.} 
 \label{flo1}
\end{figure}

To incorporate radial flow in the model, we modify the phase
space distribution (Eqs. \ref{1},\ref{2}) by introducing
a radial flow $\beta(r)$ of the emission points,  
  \begin{equation}
    \beta(r) = 1 - e^{-r/f}\, ,  
    \label{4}
  \end{equation}
where  $f$ is an adjustable parameter. For different flow strengths
$f$, the radial dependence is shown in Fig.~\ref{flo1}.
Superimposed is the mean value of the radial velocity
  \begin{equation}
    \beta_{r} = \frac{\vec{p}\cdot\vec{r}}{E \mid r \mid}\, ,
    \label{3}
  \end{equation}
extracted from simulated Pb+Pb Venus events, version 4.12 \cite{Wer1}.
Here $E$ denotes the total energy of the pion and $r$ its radial
distance from the source center. 
One sees that a choice of $f=9$ fm fits the Venus data very well.
This value of the flow strength, $f=9$,
was used for simulating a ``realistic'' flow.
No attempt was made to reproduce any other Venus distribution. 
As demonstrated in \cite{Wie1}, 
the flow dependence of the observables is mainly due
to the size of the flow, while its functional shape plays
a somewhat secondary role. The order of magnitude of the flow 
velocity extracted here ($\sim 0.35c$) 
can be compared to the flow parameter $\eta_f$ in \cite{Wie1}.

\begin{figure}[p]
 \begin{center}\mbox{\epsfysize=7cm\epsffile{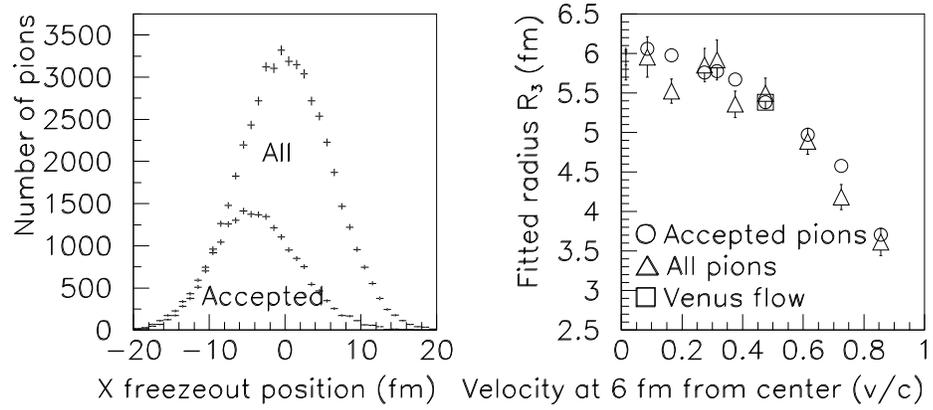}}\end{center}
 \caption{{\bf Left}: Freezeout positions in the $x$ direction for all
pions and those satisfying a kinematical acceptance cut, 
where the initial pion 
momenta have been changed by adding radial flow extracted from 
Venus (eg. Eq. (\ref{7}) with $f$=9 fm)). 
{\bf Right}: The HBT radius $R_3$ as a function of the flow velocity 
at a distance of $r$=6 fm from the source center
for the radial flow profiles in Figure ~\ref{flo1}.} 
 \label{r3flow1}
\end{figure}

 We have generated events for different values of the flow 
strength $f$ and the result was verified to be independent of statistics. 
Each Monte-Carlo event contained typically 100 identical pions. For the 
relative pair momentum differences, 5 MeV bins 
were used and the results were checked to be insensitive
to bin sizes in the range of 5-20 MeV. The original aim was
the study of the flow and acceptance dependence of HBT radius
parameters. Indeed, our simple model shows reasonable 
physical properties. Especially the flow dependence of the 
1- and 3-dimensional $k_T$ integrated HBT radius
parameters is in qualitative agreement with that obtained in other
model studies,\cite{Pra3,Wie1,Fie1}: 
The HBT radii decrease with increasing flow strength $f$,
since the effective emission region (``region of homogeneity''~\cite{Akk1})
for pion pairs with small relative momenta decreases for increasing $f$. 
Also, we have considered the so far little studied effect of the detector 
acceptance on the HBT radius parameters.
In Figure ~\ref{r3flow1} (left), we show
for the case of a realistic flow strength $f = 9$ fm
the freezeout positions of all generated pions and of those
satisfying the ($p_T,y$) acceptance criteria
of a typical magnetic spectrometer~\cite{Ros1}.
The emission region of the detected pions is clearly smaller than
the total emission region.
The corresponding HBT radius parameter $R_3$ 
is presented in Figure ~\ref{r3flow1} (right) as a function of the flow 
strength $f$ for the cases with a realistic magnetic spectrometer 
acceptance and without. In both cases, the HBT radii decrease with
increasing flow and the acceptance dependence is very small.
>From all this we conclude here only that our model is
not too oversimplified and reproduces essential features obtained
in more complete simulations.  

We now consider a simple modification of our model, the
introduction of an absorption cut. To this aim, we impose
on the Monte Carlo output, defined by the Eqs. (\ref{1}-\ref{3}),
a kinematical cut which effectively strengthens the position-momentum
correlation in the source: pions are only emitted, if their
momentum vector is in an angle less than 45 degrees around 
the direction of their position vector, i.e., each pion has to move 
out of the source in a 45 degree cone.
Under these conditions, the correlation functions in 
$q_3$ displayed in Figure ~\ref{flopt1} have been obtained,
where, additionally, cuts in single particle transverse momentum 
of 0.1 GeV/c - 0.3 GeV/c have been applied. 

We emphasize that in contrast to the pathological source discussed in
subsection \ref{sec3.1}, this modified source shows a rather reasonable
phase space distribution which is difficult to reject a priori. Still,
the correlator obtained from the $\cos$-prescription (Eq. \ref{5}) 
again drops below unity. Especially, when $p_T$ is restricted to 
small values, the dip in the correlation function around 0.08 GeV/c 
figures more prominently. Such a characteristic dip 
can be produced in $C(q_3)$ given at least a 
$p_T$ cut, and {\em either} radial flow {\em or} the angle cut.
We thus conclude that it arises when imposing an acceptance cut 
on the pions emitted from 
the Gaussian model with strong position-momentum correlation.
It is worth noting that the half-width of the correlation 
function remains nearly identical in all four cases in Figure ~\ref{flopt1}.

\section{Discussion}

Existing event generators do not propagate (anti)-symmetrized wave 
functions and hence face a conceptual
difficulty in incorporating the effect of Bose-Einstein 
correlations ~\cite{Aic1}. The current practice of modifying the
weight of pion pairs by the Born probability $\mid \psi \mid^2$ of
symmetrized plane waves does not address this problem properly.
Here, we have shown for the first time that this conceptual
difficulty can have significant quantitative consequences. In fact,
the dip observed in the correlator Fig.~\ref{flopt1} is clearly
unphysical and shows that the formalism leading to Eq. (\ref{6}) 
breaks down. This becomes clear if one looks at the formalism
used for analytical models of the emission function, where the
ad hoc prescription (\ref{5}) is not needed. In these, the
evaluation of the Bose-Einstein correlation function is based 
on the coherent state formalism arising from
quantum field theory, using \cite{Hei1},\cite{Wie1}
  \begin{equation}
     C({\bf q},{\bf K}) = 1 + 
     {\left\vert \int d^4x\, S(x,K)\, e^{iq{\cdot}x}\right\vert^2 
      \over
      \int d^4x\, S(x,K+\frac{1}{2}q) \int d^4x\, S(x,K-\frac{1}{2}q)}
    \label{10}
  \end{equation}
with $q = p_1 - p_2$ 
and $K = (p_1 + p_2)/2$. Clearly, this 
correlation function cannot become smaller than unity. The approximate
method using Eq. (\ref{6}) is based on a semiclassical picture
for a set of discrete space-time points. It produces effects that
are not always consistent with Eq. (\ref{10}) and which can become 
non-negligible as seen in Figure ~\ref{flopt1}. They are more
pronounced in the presence of strong position-momentum dependence
when the long-range characteristics of the argument
$({\vec{r_1}}-{\vec{r_2}}) \cdot ({\vec{p_1}}-{\vec{p_2}})$
in (\ref{6}) play an important role.

The discrepancy in methodology can be even more drastic; an analytical
calculation based on Eq. (\ref{10}) using the source given
in Eq. (\ref{8}) does not show these oscillations \cite{Hei2}, while
using Eq. (\ref{6}) with the same source yields Figure ~\ref{zpz1}.
However, Eq. (\ref{6}) is at this time the only method available to 
build correlation functions from the space-time
output of microscopic event generators~\cite{Pra2},\cite{Sul1},
\cite{Hum1},\cite{Hum2},\cite{Fie1}. This formalism has yielded 
reasonable results under the less severe conditions
studied until now. The exact expression given in Eq. (\ref{10}) however 
cannot be used in a Monte-Carlo simulation when the Monte-Carlo event 
generator does not provide the source density function of the mean 
momentum ${\bf K}$. The probabilistic Monte-Carlo approach does not allow 
to deal with the quantum mechanics effects involved here. Without a solution
to this problem, since the procedure based on Eq. (\ref{6}) has been used 
by many groups in the recent literature 
\cite{Pra2},\cite{Sul1},\cite{Hum1},\cite{Fie1}, it is now henceforth 
clearly important to improve the current Monte-Carlo Bose-Einstein formalism 
and to estimate the errors involved in the procedure using Eq. (\ref{6}).

\section{Conclusions}

We have shown that the $\cos$-prescription commonly used
to include Bose-Einstein effects in the Monte-Carlo simulation
can lead to results not consistent with first principles.
This calls into question its quantitative and qualitative reliability, 
especially for the case that certain kinematical selections are applied when 
strong position-momentum correlations are inherent in the pion source. 
A deeper understanding of this simulation formalism is necessary 
in order to make more detailed analyses of dynamical issues and 
acceptance effects. Most remarkably, the $\cos$-prescription 
(\ref{5}) interprets both position $\vec{r}_i$, $t_i$, and momenta
$\vec{p}_i$ returned by an event generator as sharp (classical)
phase space coordinates. This clearly violates the Heisenberg uncertainty
principle. On the other hand, a smearing of the emission points, 
motivated e.g. by the picture of a limited quantum mechanical
particle localization, allows to remedy the unphysical dip in the
correlator at least partly, see Section~\ref{sec3.1}. This
observation may indicate 
in our opinion that the inconsistencies of the prescription (\ref{5}) 
presented here are rooted in an incorrect treatment of the quantum 
mechanical particle localization.  
This points out the need for an advanced quantum 
mechanical Monte-Carlo event generator that can properly describe 
Bose-Einstein correlation functions.

\section{Acknowledgements}

\begin{figure}[p]
 \begin{center}\mbox{\epsfysize=10cm\epsffile{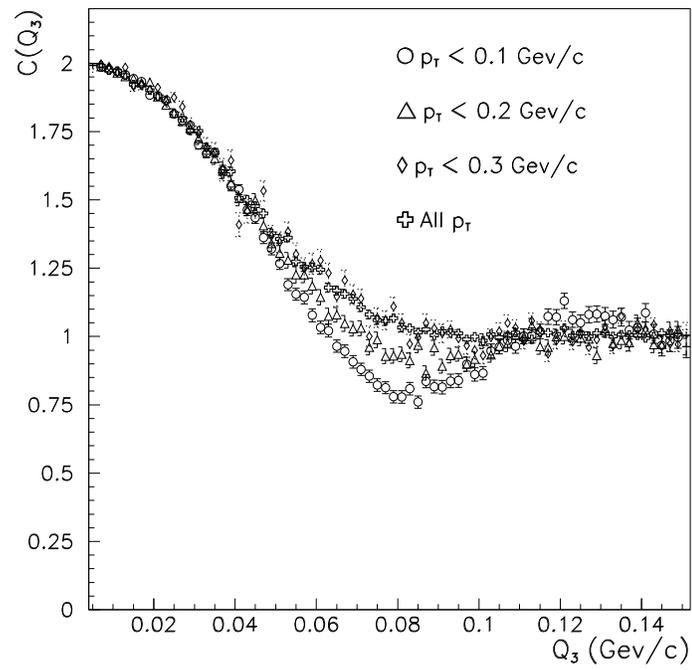}}\end{center}
 \caption{Pion correlation in $q_{3}$, with a horizon cut imposed such
that the pion momentum vector must be within a 45 degree angle of the radial
vector, for four different selections in single pion transverse momentum.}
 \label{flopt1}
\end{figure}

 The authors gratefully acknowledge the active contribution of
U. Heinz, in particular the critical reading 
of this manuscript, and H.-P. Naef, who brought attention to
the behaviour shown in Figure ~\ref{flopt1}. This work has 
been supported by the {\em Fonds national suisse de la recherche scientifique} 
under contract 20-43'126.95.

\end{document}